\documentclass[superscriptaddress,preprint,preprintnumbers,amsmath,amssymb]{revtex4}
\usepackage{graphicx,afterpage}
\usepackage{txfonts}% Physical Review B
\usepackage{dcolumn}% Align table columns on decimal point
\usepackage{bm}% bold math
\usepackage{array}
\begin{document}

\preprint{APS/123-QED}
\title{Electron mass enhancement %of the %zero-temperature London penetration depth %
near a nematic quantum critical point in %the electron-doped
 NaFe$_{1-x}$Co$_{x}$As}
\author{C. G. Wang}
\affiliation{Institute of Physics, Chinese Academy of Sciences,\\
	and Beijing National Laboratory for Condensed Matter Physics,Beijing 100190, China}
\affiliation{School of Physical Sciences, University of Chinese Academy of Sciences, Beijing 100190, China}
\author{Z. Li}
\affiliation{Institute of Physics, Chinese Academy of Sciences,\\
	and Beijing National Laboratory for Condensed Matter Physics,Beijing 100190, China}
\affiliation{School of Physical Sciences, University of Chinese Academy of Sciences, Beijing 100190, China}
\author{J. Yang}
\author{L. Y. Xing}
\affiliation{Institute of Physics, Chinese Academy of Sciences,\\
	and Beijing National Laboratory for Condensed Matter Physics,Beijing 100190, China}
\author{G. Y. Dai}
\affiliation{Institute of Physics, Chinese Academy of Sciences,\\
	and Beijing National Laboratory for Condensed Matter Physics,Beijing 100190, China}
\affiliation{School of Physical Sciences, University of Chinese Academy of Sciences, Beijing 100190, China}
\author{X. C. Wang}
\affiliation{Institute of Physics, Chinese Academy of Sciences,\\
	and Beijing National Laboratory for Condensed Matter Physics,Beijing 100190, China}
\author{C. Q. Jin}
\affiliation{Institute of Physics, Chinese Academy of Sciences,\\
 and Beijing National Laboratory for Condensed Matter Physics,Beijing 100190, China}
\affiliation{School of Physical Sciences, University of Chinese Academy of Sciences, Beijing 100190, China}
\author{R. Zhou}
\affiliation{Institute of Physics, Chinese Academy of Sciences,\\
	and Beijing National Laboratory for Condensed Matter Physics,Beijing 100190, China}
\author{Guo-qing Zheng}
\affiliation{Institute of Physics, Chinese Academy of Sciences,\\
 and Beijing National Laboratory for Condensed Matter Physics,Beijing 100190, China}
\affiliation{School of Physical Sciences, University of Chinese Academy of Sciences, Beijing 100190, China}
 \affiliation{Department of Physics, Okayama University, Okayama 700-8530, Japan}
\date{\today}% It is always \today, today,
             %  but any date may be explicitly specified

\begin{abstract}
{
{ A magnetic order can be completely suppressed at zero temperature ($T$), by doping carriers or applying  pressure,  at a quantum critical point (QCP) , around which physical properties
	change drastically. %, which can be probed by  the zero-temperature London penetration depth $\lambda_{\rm L}$(0).
	%As a local probe,
	However, the situation is unclear for an electronic nematic order that breaks rotation symmetry.
	Here we report % $^{75}$As and $^{23}$Na
	nuclear magnetic resonance (NMR) studies on NaFe$_{1-x}$Co$_{x}$As where  magnetic and nematic transitions %occur. % but the two transition temperatures
	are well separated.
	The NMR spectrum is sensitive to inhomogeneous magnetic fields   %$\Delta B$
	%created by the vortices
	in the vortex %superconducting
	state, which is  related to  London penetration depth $\lambda_{\rm L}$ that measures the electron mass $m^*$. We discovered two  peaks in the doping dependence of %the zero-temperature-limit
	$\lambda_{\rm L}^2$($T\sim$0); one at %the magnetic QCP
	$x_{\rm M}$=0.027  where  the spin-lattice relaxation rate shows quantum critical behavior, and another  at $x_{\rm c}$=0.032 around which the  nematic transition temperature extrapolates to zero and
	the electrical resistivity shows a $T$-linear variation. % revealed by the spin-lattice relaxation rate measurement in the normal state.
	Our results indicate that a nematic QCP lies beneath the superconducting dome at $x_{\rm c}$ %= 0.032
	where $m^*$  is enhanced. The impact of the nematic fluctuations on superconductivity is  discussed.} %, and %which suggest an important role of quantum nematic fluctuations on superconductivity.
	%various properties and  superconductivity itself.} %a possible link  between  iron-based and cuprate high-temperature superconductors.} %suggest the ubiquitousness of nematic states in quantum materials.} %
%ng pairing interaction.}
}
\end{abstract}

%We reveal that the optimal composition \emph{x}=0.027 with $T_{c}$=20.5 K is a magnetic Quantum Critical Point (QCP). By detecting the distribution of
%the magnetic field which we obtained from the broaden of the $^{23}$Na center peak, we further invest the doping \emph{x} evolution the Magnetic Penetration Depth \textrm{$\lambda$}. We find the peak of \textrm{$\lambda$} to be at the slightly overdoped composition
%\emph{x}=0.032. This result is different with the former report on BaFe$_{2}$(As$_{1-x}$P$_{x}$)$_{2}$ in which \textrm{$\lambda$} meets a sharp peak at the optimal composition and it is considered to be related to the magnetic QCP.


\maketitle

In the  high transition-temperature ($T_{\rm c}$) superconducting cuprates or iron pnictides, superconductivity adjoins a magnetically-ordered % or/and an electronic nematic ordered
phase \cite{Cuprate_review,Fernandes_review}. With increasing carrier doping or externally-applied pressure to a parent phase,  the magnetic %and electronic nematic
order is suppressed and a superconducting phase emerges. The magnetic order temperature $T_{\rm N}$ goes to zero before  superconductivity appears or extrapolates to zero at a point  inside a superconducting dome. % \cite{Keimer_cuprate}. %Chubukov_1
Around the  ending point of $T_{\rm N}$ = 0, namely, a quantum critical point (QCP), many anomalous %normal-state
physical properties due to the associated quantum fluctuations have been revealed by various experimental methods  \cite{Cooper,HashimotoScience,Chu2,ZhouNatCommun}.
A magnetic QCP is considered by many a key to understand the  mechanism of  high-$T_{\rm c}$ superconductivity \cite{Coleman_nat}.
For example, the electron pairing strength is believed to be enhanced by the magnetic quantum fluctuations \cite{pairing}.

In iron pnictides, in addition to the magnetic order, there also exists an electronic nematic order that breaks rotation symmetry, setting in at the Tetragonal-to-Orthorhombic structural transition temperature $T_{\rm s}$ or even above \cite{Chu2,Xly,S4,ZhouPRB}, which has attracted much attention recently. % \cite{Lederer_PNAS}.
It was proposed that such nematic order may stem from the electronic orbital
degree of freedom, in addition to spin degree of freedom  \cite{Fernandes_1,Kontani,Kuo_SCI1}. Thus the electronic nematicity points to a new frontier of condensed matter physics \cite{Chubukov_1,Schmalian2014} and may also hints at a possible new route to high-$T_{\rm c}$ superconductivity \cite{Yang,Lederer_2,Lee}.
Although some anomalous  physical properties such as temperature ($T$)-linear electrical resistivity or diverging nematic susceptibility behavior can be understood as due to nematic quantum fluctuations at high temperatures \cite{ZhouNatCommun,Lederer_PNAS,Kuo_SCI1}, a direct evidence for a nematic QCP  inside the superconducting dome is still lacking. %,
%as superconductivity sets in at a high temperature in many cases,
 % \cite{Brun}.

If a QCP is indeed hidden inside the  dome, it would manifest itself in some physical quantities that describe the  zero-$T$-limit properties. London penetration depth $\lambda_{\rm L}$ is determined by the superfluid density $n$ and the effective mass $m^{*}$ of carriers responsible for superconductivity \cite{Varma_1}, and $\lambda_{\rm L}$($T$ = 0) can be a good tool for probing a hidden QCP. %, which is an important physical quantity to study electron state in superconducting state.
This is because many experiments indicated that $m^{*}$ can be enhanced %upon approaching a QCP
due to %the increased
quantum fluctuations \cite{Ramshaw,HashimotoScience,Walmsley}.
In the cuprate superconductor YBa$_2$Cu$_3$O$_{6+\delta}$, as the magnetic QCP is approached from the underdoped side,   $m^{*}$ increased  by a factor of 3 \cite{Ramshaw}.
In the isovalent-doped Fe-based superconductor BaFe$_{2}$(As$_{1-x}$P$_{x}$)$_{2}$,
%In fact, previous measurements in the isovalent doped system BaFe$_{2}$(As$_{1-x}$P$_{x}$)$_{2}$ found
a sharp peak of  $\lambda_{\rm L}$(0) was indeed found at the optimal doping concentration $x$ = 0.3, which was attributed to an antiferromagnetic   QCP \cite{HashimotoScience}.
Quantum oscillation measurements confirmed that upon decreasing  $x$ from 0.8 to 0.3, $m^{*}$  is doubled \cite{Walmsley}.

 %In electron-doped system, two kinds of phase transitions are more separated\cite{Chubukov_1}.
%However, early study on Ba(Fe$_{1-x}$Co$_{x}$)$_{2}$As$_{2}$ did not find any special feature of doping dependent $\lambda_{\rm L}$, maybe due to greater degree of electronic disorder caused by Co doping or different superconducting pairing symmetry\cite{Luan,Gordon}.
%%So it is still unclear whether a peak of $\lambda_{\rm L}$ could also exist in a clean carrier-doped system, and which kind of quantum fluctuation should be responsible for it.

%Takuya Nomoto et al\cite{Nomoto} consider that the critical magnetic fluctuation renormalizes the current vertex and drastically enhances the zero-temperature penetration
%depth. However, in the BaFe$_{2}$(As$_{1-x}$P$_{x}$)$_{2}$ system, a tetragonalto-orthorhombic structural transition temperature $T_{s}$ coincides with the magnetic transition temperature $T_{N}$\cite{ZhouNatCommun} and the structure phase transition line also crosses into the SC dome and the extrapolated structure QCP is quite close to the magnetic one, making it difficult to clarify the relationship between the magnetic fluctuation and the observed peak of the penetration depth \textrm{$\lambda$}.

In the iron-pnictides, the putative magntic QCP and electronic nematic QCP are usually close-by or even indistinguishable, which hindered the progress of experimental investigations on the later.
NaFe$_{1-x}$Co$_{x}$As is an exceptional system \cite{NaFeAs} whose   $T_{\rm s}$ is  about 10 K higher than  $T_{\rm N}$ in the parent undoped compound. With Co doping, $T_{\rm N}$  is suppressed much more rapidly than $T_{\rm s}$. The difference between the two transitions increases to   20 K at $x$ = 0.018 \cite{ZhouPRB}. In the orthorhombic phase, electronic nematicity was visualized  in the parent compound by %a local probe,
scanning tunneling spectroscopy \cite{Xly}. Subsequently, both orbital %order
and spin nematicity were observed above $T_{\rm s}$  %orthorhombic state
not only in parent compound but also in doped samples by nuclear magnetic resonance (NMR) \cite{ZhouPRB}.
%particularly suitable electron-doped system to study the connection between high-$T_c$ superconductivity and different QCP.

The  Co-doping concentration to  obtain the highest $T_{\rm c}$ is only $\sim$2.7 percent that is much smaller than any other systems \cite{NaFeAs}. As  demonstrated by the much narrower $^{75}$As-NMR lines \cite{supplemental_materials}, the doping-induced disorder in the FeAs plane, which is usually harmful to a QCP, is much less   compared to other  systems.  %Secondly, $T_{N}$ is well separated from $T_{s}$, so that the two different transitions  can be well distinguished  \cite{ZhouPRB}.
These advantages  provide one a unique opportunity to explore a nematic QCP and its influence on the physical properties.

In this Letter, %we combine NMR and transport measurements on NaFe$_{1-x}$Co$_{x}$As ({0.0089} $\leq$ $x$ $\leq$ {0.056}).
through $^{23}$Na NMR spectrum measurements, we present a detailed study of %London penetration depth
$\lambda_{\rm L}^2$($T\sim$~0) in NaFe$_{1-x}$Co$_{x}$As ({0.0089} $\leq$ $x$ $\leq$ {0.056}). % obtained from
%the distribution of the magnetic field in the superconducting state.
We find two peaks in the doping dependence of $\lambda_{\rm L}^2$($T\sim$~0); one at $x_{\rm M}$ = 0.027, %which is a magnetic QCP as evidenced by $^{75}$As  spin-lattice relaxation,
and the other at $x_{\rm c}$ = 0.032. %, at which the nematic transition temperature extrapolates to zero and the electrical resistivity shows a $T$-linear variation.
Our results provide compelling evidence that a   nematic QCP lies beneath the superconducting dome at $x_{\rm c}$ where $m^*$ is enhanced. %, and suggest  the important role of quantum nematic fluctuations  in iron-based superconductors.

The single crystals of NaFe$_{1-x}$Co$_{x}$As used in this study were grown by the self-flux method \cite{Xly}. The Co content $x$ was determined by energy-dispersive x-ray spectroscopy, and checked by the doping dependence of $^{23}$Na-NMR Knight shift (see  Fig. S1 \cite{supplemental_materials}). In order to prevent sample degradation, the samples were covered by epoxy (Stycast 1266) in a glove box filled with high-purity Ar gas before %NMR
 measurements. The typical sample size is 2mm$\times$2mm$\times$0.2mm. The  $T_{\rm c}$ was determined by DC susceptibility measured by using a superconducting quantum interference device. The $^{23}$Na NMR spectra were obtained by Fast Fourier Transform (FFT) of the spin echo and the $^{75}$As NMR spectra were obtained by integrating the spin echo as a function of frequency. The $T_1$ was measured by using the saturation-recovery method, and determined by a good fitting of the nuclear magnetization to $1-\frac{M\left( t \right)}{M\left( \infty  \right)}=0.9{{\exp }^{-6t/{{T}_{1}}\;}}+0.1{{\exp }^{-t/{{T}_{1}}\;}}$, where $M(t)$ is the nuclear magnetization at time $t$ after the saturation pulse. %All the measurements were performed at $B_{0}$ = 12 T. The resistivity was measured by the standard DC four-probe method in a Physical Properties Measurement System (Quantum Design). The electrodes were made in a glove box filled with high-purity Ar gas and the samples were covered with N grease before measurements to prevent sample degradation. The temperature varying rate during the measurements is 1 K/minute.
\begin{figure}
\includegraphics[width=0.48\textwidth]{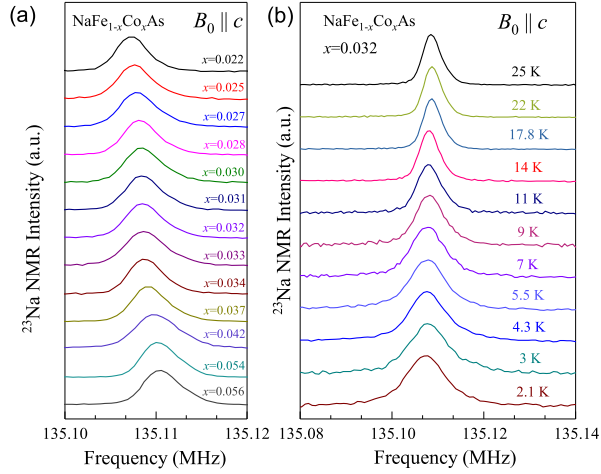}
\centering
\caption{(a)	The $^{23}$Na spectra for 0.022 $<$ \emph{x} $<$ 0.056 in the normal state ($T$ = 25 K) at $B_0$ = 12 T.
 (b) Typical temperature evolution of the $^{23}$Na spectra for the \emph{x} = 0.032 sample at $B_0$ = 12 T.
}
\label{spectra}
\end{figure}

In the vortex state, the magnetic field ${{B}_{0}}$ penetrates into a sample in the unit of quantized  flux ${{\phi }_{0}}=2.07\times {{10}^{-15}}\text{T}\cdot {{\text{m}}^{2}}$. %For a given magnetic field $B_0$,
Since the vortices  form a triangular or square lattice, the magnetic field becomes inhomogeneous in a sample. % \cite{ZhengPRL}. % below irreversible temperature $T_{irr}$.
For ${{B}_{\rm c1}}\ll {{B}_{0}}\ll {{B}_{\rm c2}}$, where $B_{\rm c1}$ and $B_{\rm c2}$ are  the  lower and upper critical field, respectively.
 The field distribution $\triangle B$ can be written as \cite{Brandt}

\begin{equation}
\Delta B=0.0609\frac{{{\phi }_{0}}}{\lambda _{L}^{2}},
\label{dB}
\end{equation}
%Such a magnetic field distribution The $\triangle B$
which can be detected by the $^{23}$Na- or $^{75}$As- NMR spectrum broadening  $\Delta$$f$ = $\gamma$$_{n}$$\triangle$$B$, %in NaFe$_{1-x}$Co$_{x}$As,
where $\gamma_n$ is the gyromagnetic ratio and $B_{c1} < $ 0.005 T and $B_{c2} >$ 44 T for  0.02 $<$ $x$ $<$ 0.05 \cite{Ghannadzadeh}.
%However,
The $^{23}$Na nuclear spin has a  larger $\gamma$$_{n}$ than $^{75}$As, making it a better probe for
$\triangle$$B$. In addition,
%$^{23}$Na is far away from the FeAs layer, and thus is insensitive to the electronic excitations or inhomogeneity. As a result,
the $^{23}$Na-NMR central  ($1/2\leftarrow \rightarrow$-1/2) transition line is much narrower than that of  $^{75}$As-NMR \cite{Halperin}. %, and is also insensitive to the inhomogeneous electronic state in the superconucting state. %, like Andreev bound state inside vortex core.
These  advantages give the $^{23}$Na-NMR spectroscopy a  high resolution for determining  $\lambda _{\rm L}$.

%NaFe$_{1-x}$Co$_{x}$As is a type-superconductor and it comes into a mixed state if the applied magnetic field $H_{0}$ meets the conditions $ H_{c1}$$<$$H_{0}$$<$$H_{c2}$. Considering FWHM($T$)-FWHM($T$$>$$T_{c}$)=$\gamma$$_{n}$*$\triangle$$B$, we have
%$\lambda$($T$)$^{2}$=0.0609$\phi$*$\gamma$$_{n}$/ [FWHM($T$)-FWHM($T$$>$ $T_{c}$) ](equation 1). Here FWHM is the full width of the NMR spectrum at half maximum.

Figure \ref{spectra} (a) shows  the $^{23}$Na-NMR spectra for various samples, and Fig. \ref{spectra} (b) shows
the $T$-dependence of the  spectrum for $x$ = 0.032. In the normal state, the spectrum is  well fitted by a single Lorentz function with a  full width at half maximum (FWHM) of $\sim$~4 kHz at $B_0$ = 12 T. The almost same width is obtained for all samples with different $x$ \cite{supplemental_materials}, which indicates a high sample quality. In the superconducting state, the central transition line % (1/2$\leftarrow \rightarrow$-1/2)
is broadened nearly symmetrically and can also be fitted by a Lorentzian function \cite{supplemental_materials}. Since the FWHM of a convolution of two Lorentzian functions is the sum of  individual FWHMs, the line broadening %due to the vortices
can be obtained by simply subtracting the $T$-independent width at high temperatures, $\Delta$$f$ = FWHM($T$) - FWHM($T$~$>$~ $T_{c}$)~\cite{supplemental_materials}.

\begin{figure}
\includegraphics[width=0.45\textwidth]{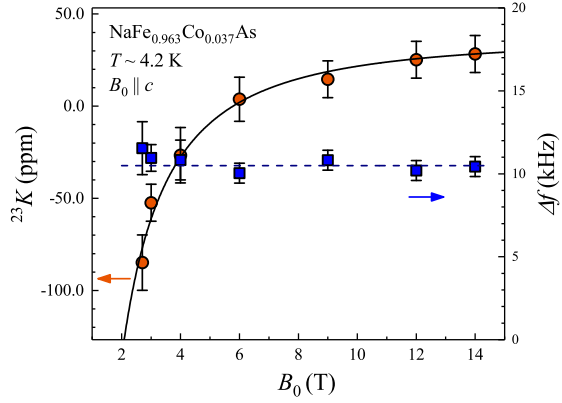}
\caption{ Field dependence of $^{23}K$ and line broadening $\Delta f$=FWHM($T$=4.2 K)-FWHM($T$=25 K) for the $x$ = 0.037 sample. Solid curve is a fitting to Eq. \ref{Kdia} by taking $D$ = 0.8. The dashed line is a guide for the eyes.}% The error bar is the s.d. in fitting the NMR spectrum.}
\label{K_dia}
\end{figure}

Theoretically, the magnetic field distribution due to the vortex lattice formation should introduce an asymmetric broadening so that a "Redfield Pattern" lineshape will be observed. In reality, however, such pattern is seldom seen in correlated systems except for  limited examples \cite{Curro,Mitrovic,ZhengTl2201,Kumagai}. In the current case, no clear "Redfield Pattern" is  observed  down to $B_0$ = 3 T. % \cite{supplemental_materials},
%even though  NaFe$_{1-x}$Co$_x$As is  clean %quasi-2D
%system and the line shape is still Lorentz at low temperatures. % which means that the vortex pinning is small so that Eq. \ref{dB} is applicable.
The symmetric line shape is  likely due to %the layered structure that causes the
flux-line  oscillations along the $c$-axis which creates a vortex lattice disorder  between different layers \cite{BrandtPRL}.  %effectively widens the vortex core and smears the magnetic field of each flux line \cite{BrandtPRL}.
Indeed, a symmetric magnetic-field distribution was observed in Bi$_2$Sr$_2$CaCu$_2$O$_6$ by $\mu$sr in the superconducting state and explained by such "disordered fluxon model" \cite{Bi2212}.
%
 % The vortex-cores may have small random displacements from triangular lattice, resulting in the broadening of effective core radius. This will truncate the high-field tail in field distribution, then line shape becomes symmetric\cite{supplemental_materials}.
% However, this effect will modify a little bit of the prefactor in Eq. \ref{dB}\cite{supplemental_materials}.

In order to experimentally demonstrate that  Eq. \ref{dB} is indeed valid, we show the  line shift and the  line broadening $\Delta$$f$ at various fields in Fig. \ref{K_dia}. A field-independent  $\Delta$$f$ is clearly seen as expected by  the London theory %that $\Delta_B$ should be independent of the field for a field higher than
for a field above $B_{c1}$, indicating that the line broadening is indeed caused by the vortices.
Another evidence for the observed NMR line broadening stemming from the vortex lattice is that the   shift is progressively reduced with decreasing field as seen in Fig. \ref{K_dia}. Such diamagnetism is a solid evidence  %which indicates that the peak experiences   expected
for vortex lattice formation.
%For $^{23}$Na-NMR, the hyperfine coupling to spin susceptibility is very small, so the decreasing below $T_c$ is entirely from diamagnetism(see supplemental materials for $x$ = 0.032)\cite{supplemental_materials}.
 The diamagnetic shift $^{23}K_{dia}(B_0)$ % = ^{23}K(T, B_0) - ^{23}K(T = T_c, B+0)$
is also related to $\lambda_{\rm L}$  as~\cite{Gennes},
%Another important feature of NMR lines in the vortex state is that the line will shift to lower frequency with decreasing temperature due to diamagnetism. For $^{23}$Na-NMR, the hyperfine coupling to spin susceptibility is very small, so the decreasing below $T_c$ is entirely from diamagnetism(see supplemental materials for $x$ = 0.032)\cite{supplemental_materials}. The diamagnetic shift $^{23}K_{dia}(B_0) = ^{23}K(T, B_0) - ^{23}K(T = T_c, B+0)$ is also related to the penetration depth as~\cite{Gennes},
%
\begin{equation}
{}^{23}K={}^{23}{{K}_{0}}+{}^{23}{{K}_{\text{dia}}}\left( {{B}_{0}} \right)={}^{23}{{K}_{0}}-\left( 1-D \right)\frac{{{\phi }_{0}}}{8\pi {{\lambda_L }^{2}}{{B}_{0}}}\ln \left( \frac{4\pi\beta^2}{e\sqrt{3}}\frac{{{B}_{c2}}}{{{B}_{0}}} \right)
%^{23}K = ^{23}K_0 + ^{23}K_{\rm{dia}}(B_0) \\
%=^{23}K_0+\left( 1-D \right){{B}_{c1}}\frac{\ln \left( \frac{\beta d}{\sqrt{e}\xi } \right)}{\ln \frac{\lambda }{\zeta }} = ^{23}K_0-\left( 1-D \right)\frac{{{\phi }_{0}}}{8\pi {{\lambda }^{2}}{{B}_{0}}}\ln \left( 0.387\frac{{{B}_{c2}}}{{{B}_{0}}} \right)
\label{Kdia}
\end{equation}
%={}^{23}{{K}_{0}}+\left( 1-D \right){{B}_{c1}}\frac{\ln \left( \frac{\beta d}{\sqrt{e}\xi } \right)}{\ln \frac{\lambda }{\xi }}
%
where %, $K_0$ is the Knight shift in the normal state
$D$ is the demagnetization factor depending on the sample shape which is 0.8 for $x$ = 0.037 \cite{Pardo}, and $\beta$ is 0.38 for triangular lattice. %, $\xi$ is the superconducting coherence length and $d$ is the vortex distance.
%The crystal dimension gives $D$ =  0.8 for $x$ = 0.037 \cite{Pardo}.
It has been shown previously that $^{23}K$ is temperature independent below 100 K   although  $^{75}K$ is strongly temperature dependent \cite{Halperin},  which are confirmed by our measurements. This result indicates that the contribution from spin susceptibility to $^{23}K$  is negligible.
%, $^{23}K_0$ comes totally from orbital susceptibility which is $T$-independent.
Then we fitted the result of $^{23}K$  % field dependent $K$ at $T$ = 4.2 K
to Eq. \ref{Kdia},  as shown in Fig. \ref{K_dia}, and obtained  $\lambda_{\rm L}$  = 0.35 $\pm$ 0.03 $\rm \mu$m and $B_{c2}$ = 60 $\pm$ 20 T. Such obtained  $\lambda_{\rm L}$  is  in fair agreement with $\lambda_L(0) = 0.367$$\rm \mu$m obtained from Eq. \ref{dB} (see below). %, indicating that this is a proper equation to be used for our studies.
The deduced $B_{c2}$ is also consistent with previous report  of  $B_{c2}>$ 44 T \cite{Ghannadzadeh}. For $x$=0.03 at $B_0$=4 T, we have also confirmed that $^{23}K$ becomes negative ($\sim$-30 ppm) \cite{supplemental_materials}. % and confirmed that $^{23}K$ is negative ($\sim$-30 ppm) at that field. %Furthermore, $\delta f$ is field independent (see below).
All these assure that Eq. \ref{dB} is applicable.
%\begin{equation}
%{{K}_{\text{dia}}}\left( {{B}_{0}} \right)=\left( 1-D %\right){{B}_{c1}}\frac{\ln \left( \frac{\beta d}{\sqrt{e}\xi } \right)}{\ln \frac{\lambda }{\zeta }}=-\left( 1-D \right)\frac{{{\phi }_{0}}}{8\pi {{\lambda }^{2}}{{B}_{0}}}\ln \left( 0.387\frac{{{B}_{c2}}}{{{B}_{0}}} \right)
%\label{Kdia}
%\end{equation}

%here, $D$ is demagnetization factor, $\beta$ is 0.38 for triangular lattice of vortex, $\xi$ is the superconducting coherence length and $d$ is the average distance between vortices. The crystal dimension gives $D$ =  0.8 for $x$ = 0.037\cite{Pardo}. Then we fitted field dependent $^{23}K_{dia}$ at $T$ = 4.2K by Eq. \ref{Kdia}  as shown in Fig. \ref{K_dia}. Both penetration depth $\lambda_L$ = 0.35 $\pm$ 0.03 $\rm \mu$m and $B_{c2}$ = 60 $\pm$ 20T are deduced. This is in good agreement with $\lambda_L = 0.367$$\rm \mu$m obtained by Eq. \ref{dB}, indicating that this is a proper equation to be used for our studies. This is also consistent with previous study which shows $B_{c2}$ is higher than 40T\cite{Ghannadzadeh}.

\begin{figure}
	\includegraphics[width=0.48\textwidth]{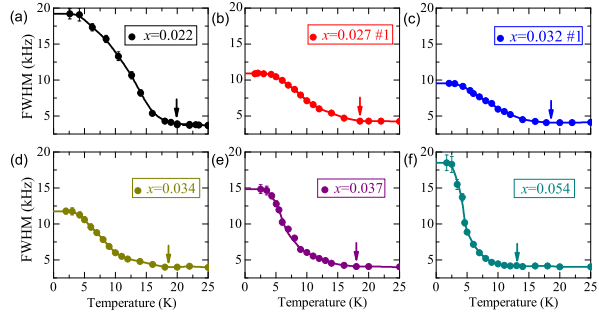}
	\caption{Temperature dependence of the full width at half maximum (FWHM ) of the $^{23}$Na-NMR spectra for various doping concentrations. The arrow indicates $T_{\rm c}$ %the superconducting transition temperature
		at $B_{\rm 0}$ = 12 T. } %The error bar is the s.d. in fitting the NMR spectrum.}
	\label{FWHM}
\end{figure}

\begin{figure}
	\includegraphics[width=0.4\textwidth]{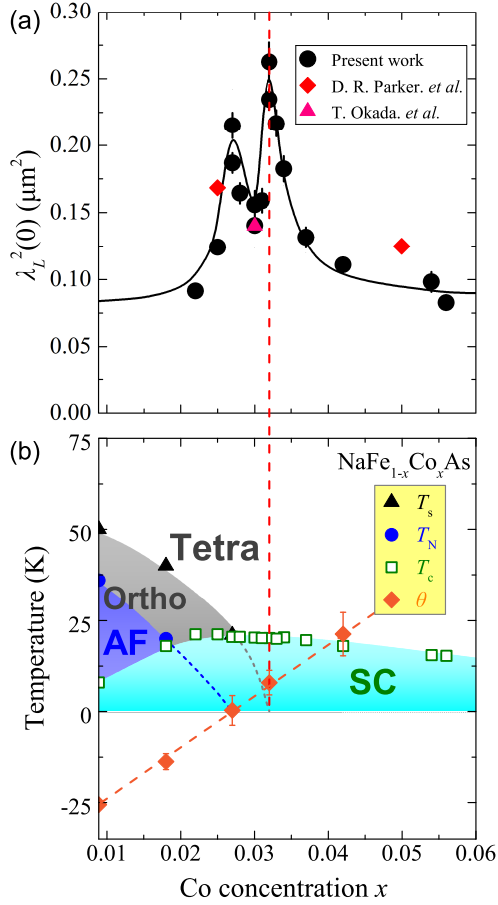}
	%\centering
	\caption{(a) $x$ dependence of the squared  London penetration depth,
		$\lambda _{L}^{2}\left( 0 \right)$. %, for various  doping concentrations.
		For $x$ = 0.027, 0.03 and 0.032, two samples were measured. The sample indicated by $\#$1 in Fig. \ref{FWHM} corresponds to a  larger  $\lambda _{L}^{2}\left( 0 \right)$. % among two samples with the same $x$.
		The red diamonds and triangle are from previous reports by other methods \cite{Parker_XRD,Okada}. %An enhancement of $\lambda _{L}^{2}\left( 0 \right)$ is clearly seen at $x_c$ = 0.032.
		The curve is a guide to the eyes. %	The error bar is from the s.d. in fitting the NMR spectrum.
		(b) The obtained phase diagram of NaFe$_{1-x}$Co$_{x}$As. %For a complete one from $x$ = 0 up to $x$ = 0.11, see Fig. S8 \cite{supplemental_materials}.
		The $T_{\rm N}$ and $T_{\rm s}$ %and $T^{*}$
		are obtained from the previous NMR spectra \cite{ZhouPRB}. %The $T^{*}$  is a temperature below which  $1/T_1$ starts to show an anisotropy between $H// a$-axis and $H// b$-axis and the $^{75}$As spectra start to be split.
		AF and SC denote antiferromagnetic ordered   and superconducting phase, respectively. Ortho and Tetra denote the orthorhombic and tetragonal crystal structure, respectively. The parameter $\theta$ is obtained from the  1/$T_{1{\rm c}}$$T$ data (see text).} %, and  $\theta$ = 0 means that the staggered susceptibility diverges at $T$ = 0. %The error bar is the s.d. in fitting the data of 1/$T_{1{\rm c}}$$T$.
	%The vertical dotted line is a guide to the eyes.}
	\label{phasediagram}
\end{figure}

Figure \ref{FWHM} shows the $T$-dependence of the FWHM. %for all doping concentrations. %without magnetic (SDW) transition.
For all samples, the line broadening  saturates  below a temperature $T_{\rm sat}$ = 0.2$\sim$~0.4~$T_c$, indicating a fully-opened superconducting gap. %We also observed very  different temperature dependence of $\lambda _{\text{L}}^{2}$ in some samples, but this is due to the doping inhomogeneity
%, strongly suggesting that sample quality is very important  for the issue we are studying@misc{ID,
%\cite{supplemental_materials}.
%The temperature dependence of $\lambda _{\text{L}}^{2}$
This is consistent with the ARPES result for the doping levels away from SDW region \cite{Liu_1,Ge}.
Then $\lambda _{\rm L}^{2}\left( 0 \right)$ is obtained  according to Eq. \ref{dB} using the data below $T_{\rm sat}$, with the results  summarized in Fig. \ref{phasediagram} (a).
The results obtained  at $B_{0}$=4 T for $x$=0.03 and 0.037  agree well with those  at $B_0$=12 T \cite{supplemental_materials}, which again assures that Eq. 1 is valid for our case.
The three data points in the figure previously reported by % muon spin rotation
$\mu$sr \cite{Parker_XRD} and by surface impedances  \cite{Okada} measurements
% $absolute value of $\lambda _{L}^{2}\left( 0 \right)$ for $x$= 0.027
%by using Eq. \ref{dB} might not be exactly identical to the real value, but
are in good agreement with our data. %comparable to that obtained from muon spin rotation measurements \cite{Parker_XRD}.

A peak is observed in the doping dependence of  $\lambda _{\rm L}^{2}\left( 0 \right)$ at $x_{\rm M}$ = 0.027. In addition, and most remarkably, an even higher peak
% a clear enhancement of $\lambda _{L}^{2}\left( 0 \right)$
is observed at $x_{\rm c}$ = 0.032.
A possibility of mesoscopic phase separation that might be responsible for an enhancement of  $\lambda _{\rm L}^{2}\left( 0 \right)$ \cite{Chowdhury1} can be ruled out, as the NMR line width at $T$=25 K shows no anomaly at $x$ = 0.027 and 0.032 (see Fig.  S9  \cite{supplemental_materials}).

In a clean single crystal, $\lambda _{\rm L}^{2}\left( 0 \right)$ is related to the electron mass as    \cite{Varma_1}
\begin{equation}
\lambda _{\text{L}}^{-2}\left( 0 \right)={{\mu }_{0}}{{e}^{2}}\sum\limits_{i}{{{n}_{i}}/m_{i}^{*}}
\label{lambda}
\end{equation}
where $\mu_0$ is the vacuum magnetic permittivity, $e$ is the electron charge, $m_{i}^{*}$ and $n_i$ are respectively the effective mass and the  superconducting carriers density in band $i$. Therefore, a peak of $\lambda _{\rm L}^{2}\left( 0 \right)$ is an indication of strong enhancement of the effective mass $m^{*}$, as $n_i$ changes monotonously  with $x$  \cite{supplemental_materials}. %due to a QCP at $x_c$ = 0.032.
In BaFe$_2$(As$_{1-x}$P$_x$)$_2$, a peak in $\lambda _{\text{L}}^{2}\left( 0 \right)$ was  found and  attributed to the existence of a magnetic QCP  \cite{HashimotoScience}, although theoretical interpretation was controversial \cite{Chowdhury1,Nomoto,Levchenko,Chowdhury}.
As we elaborate below, the first peak indicates that a magnetic QCP  lies beneath the superconducting dome at $x_{\rm M}$ = 0.027, while the higher peak indicates that an electronic nematic QCP lies beneath the dome  at $x_{\rm c}$ = 0.032.

\begin{figure}
	\includegraphics[width=0.35\textwidth]{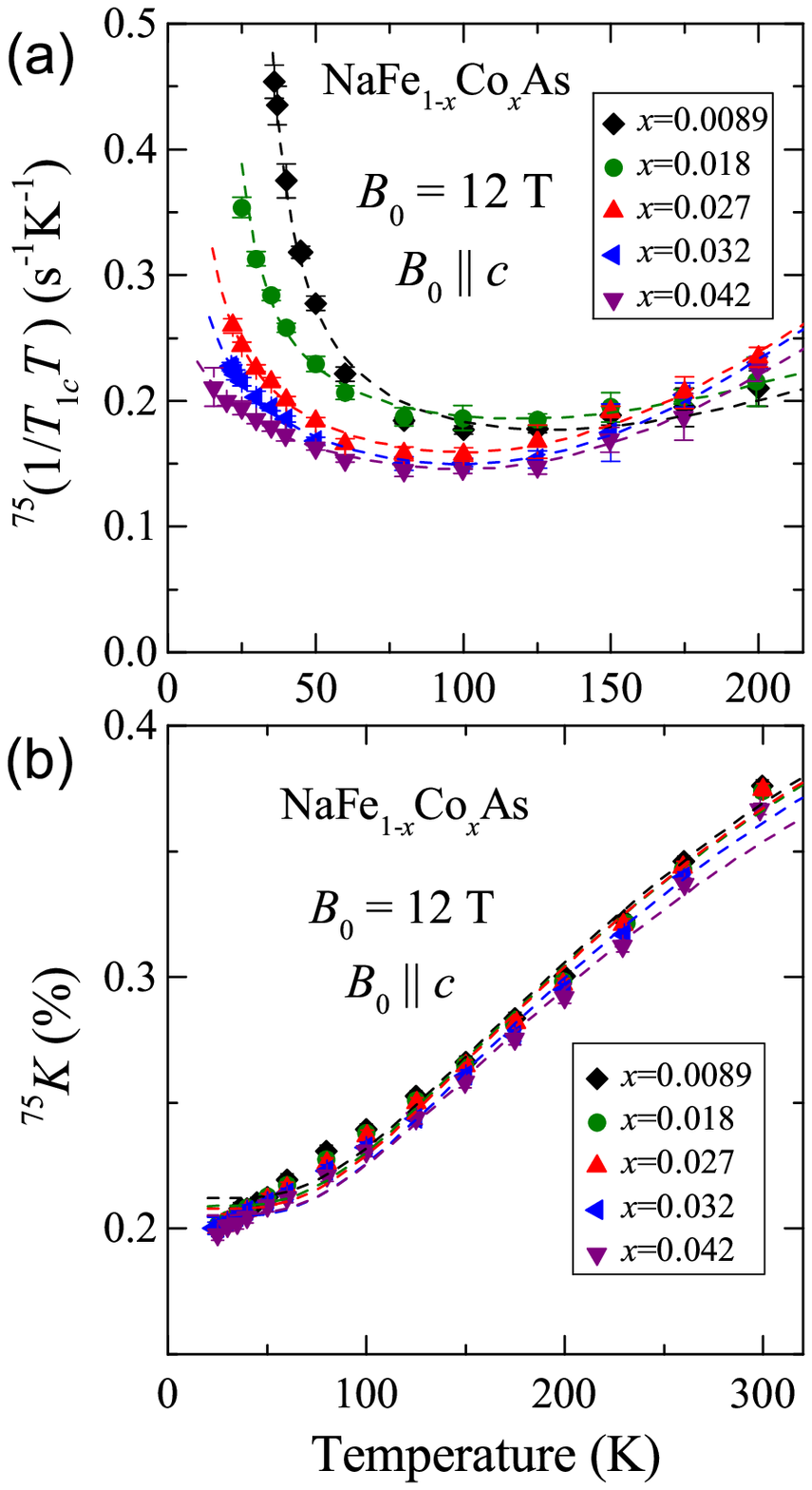}
	\caption{(a) $T$ dependence of $^{75}($1/$T_{1\rm c}$$T$) for various NaFe$_{1-x}$Co$_{x}$As. %with magnetic field parallel to the \emph{c}-axis. The error bar for $^{75}$(1/$T_{1}$$T$) is the s.d. in fitting the nuclear magnetization recovery curve.
		The dashed curves are a fit of $^{75}$(1/$T_{1}$$T$) to $^{75}$(1/$T_{1}$$T$) = \emph{a} + \emph{b}/(\emph{T}+$\theta$) + $c$$\cdot$exp(-2$E$$_g$/$k$$_B$$T$) (see text), with the obtained $\theta$  plotted in Fig. \ref{phasediagram}. (b) $T$ dependence of $^{75}$\emph{K} for various NaFe$_{1-x}$Co$_{x}$As. %The error bar for $^{75}$\emph{K} was estimated by assuming that the spectrum-peak uncertainty equals the point (frequency) interval in measuring the NMR spectra which is smaller the size of the points.
		The dashed lines are a fit of $^{75}$\emph{K} to $^{75}$\emph{K} = $^{75}$$K$$_{0}$ + $^{75}$\emph{K}$_{1}$$\cdot$exp(-$E$$_g$/$k$$_B$$T$) (see text).  }
	\label{T1T}
\end{figure}

%In the previous studies on Ba(Fe$_{1-x}$Co$_x$)$_2$As$_2$, no  peak feature in $\lambda _{\text{L}}^{2}\left( 0 \right)$ was found \cite{Luan}, probably due to a much larger  Co-doping concentration that introduced a greater degree of electronic disorder.
 %\cite{Nomoto,Chowdhury,Levchenko}. %However, it is still unclear in that system, because $T_{\rm N}$ and $T_{\rm s}$ are rather close.
%We begin by studying the spin dynamics in the normal state to locate the magnetic QCP.
%As described below, the $x_c$ = 0.032 is far away from a magnetic QCP in the present case.

\begin{figure}
\includegraphics[width=0.38\textwidth]{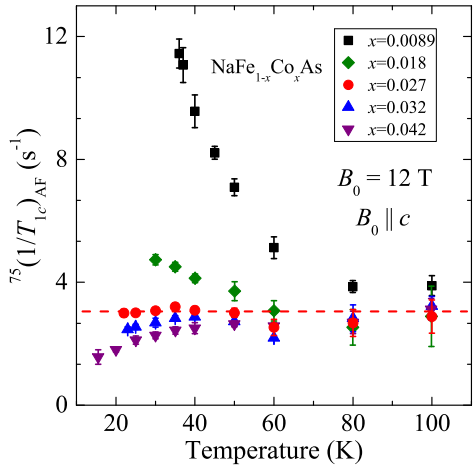}
\caption{The $T$ and $x$ dependencies of  1/$T_{1}$ due to  antiferromagnetic spin fluctuations. The $T$-independent 1/$T_{1}$ indicates a magnetic QCP.} %The error bar is obtained from the s.d. in fitting the nuclear magnetization curve and  the NMR spectrum.}
 \label{T1}
\end{figure}

 We measured the spin-lattice relaxation rate $^{75}$($1/T_{\rm 1c}$) with the  magnetic field  $B_{0}$ along the $c$-axis. % to locate the magnetic QCP. %in NaFe$_{1-x}$Co$_x$As. %As in BaFe$_{2-x}$M$_x$As$_2$ (M=Co or Ni) systems,
The quantity $^{75}$(1/$T_{\rm 1c}T$)  consists of two contributions,  $^{75}$(1/$T_{\rm 1c}T$) = $^{75}$(1/$T_{\rm 1c}T$)$_{\rm AF}$ + $^{75}$(1/$T_{\rm 1c}T$)$_{\rm intra}$, where the former  represents the contribution from antiferromagnetic spin fluctuations and the latter is from an intra-band effect \cite{ZhouNatCommun,Co-NMR-prl_Ning}. %\cite{ZhouNatCommun,Co-NMR-prl_Ning}.
The $^{75}$(1/$T_{\rm 1c}T$)$_{\rm AF}$ follows a Curie-Weiss behavior %\cite{ZhouNatCommun,Co-NMR-prl_Ning},
$b$/($T$+$\theta$), as expected for a two-dimensional itinerant electron system near a magnetic QCP \cite{itinerantelectronmagnetism}.
The intra-band contribution of $^{75}$(1/$T_{\rm 1c}T$)$_{\rm intra}$ is due to the density of states  at the Fermi level, which is related to the spin Knight shift $K_s$ according to the Korringa relation \cite{Korringa}. % $K$$_s$$^2$$(T_1T)_{\rm intra}$ = constant
As shown in Fig. \ref{T1T}~(b), the Knight shift can be fitted by $^{75}\emph{K}$ %=$^{75}K_{0}$ + $^{75}\emph{K}_{s}$
= $^{75}$$K$$_{0}$ + $^{75}$\emph{K}$_{1}$$
\times$exp(-$E$$_g$/$k$$_B$$T$), where $^{75}$$K$$_{0}$ is a constant and $^{75}$$K$$_{1}$ is $T$-dependent spin Knight shift.
%In order to obtain $\theta$ and $^{75}$(1/$T_{\rm 1c}T$)$_{\rm intra}$,
Then we can fit the 1/$T_{\rm 1c}$ data by $^{75}$(1/$T_{\rm 1c}T$) = \emph{a} + \emph{b}/(\emph{T}+$\theta$) + $c$$\times$exp(-2$E$$_g$/$k$$_B$$T$) to deduce $\theta$ as have been done in BaFe$_{2-x}$[Co,Ni]$_x$As$_2$ \cite{ZhouNatCommun,Co-NMR-prl_Ning}.  The obtained parameter $\theta$ is plotted in Fig. \ref{phasediagram}. The value of $\theta$ is almost zero for $x_{\rm M}$ = 0.027, which means that the staggered susceptibility is governed by   %$x_{\rm M}$ = 0.027 is
a magnetic QCP to become divergeing at $T$=0 \cite{itinerantelectronmagnetism}. %, indicating that it . % In fact, with $\theta$ = 0 K, $^{75}$(1/$T_{\rm 1c}T$)$_{\rm AF}$ = $a$/$T$, i.e. $^{75}$(1/$T_{\rm 1c}$)$_{\rm AF}$ =constant.
In order to see this  more visually from the $1/T_1$ data, % $^{75}$(1/$T_{\rm 1c}T$)$_{\rm intra}$ is deduced from $^{75}$(1/$T_{\rm 1c}T$) based on the fitting and the corresponding results
we plot in Fig. \ref{T1} the contribution from the antiferromagnetic spin fluctuation, $^{75}$(1/$T_{\rm 1c}$)$_{\rm AF}$, which is obtained by subtracting  $^{75}$(1/$T_{\rm 1c}T$)$_{\rm intra}$ = \emph{a} + $c$$\times$exp(-2$E$$_g$/$k$$_B$$T$) from the observed $^{75}$(1/$T_{\rm 1c}T$).
%%according to the Korringa relation  $K_{\rm s}$$^{2}$$T_{1}T$ = constant. The Knight shift $K$ has two contribution as $K$= $K_{0}$+$K_{\rm s} \cdot$exp($E_{g}$/$k_{B}T$), where $K_{0}$ is the chemical shift and is temperature independent, while the second term is due to the band effect. So we could write (1/$T_{\rm 1c}T$)$_{\rm intra}$=$b$+$c$*exp(-2$E_{g}$/$k_{B}T$). Then we fit our result by 1/$T_{\rm 1c}T$ = $a$/($T$+$\theta$)+ $b$+$c$*exp(-2$E_{g}$/$k_{B}T$)(The value of $E_{g}$ is obtained by fitting the data of Knight shift).
%After subtracting the (1/$T_{\rm 1c}T$)$_{\rm intra}$ part, the temperature dependence of (1/$T_{\rm 1c}$)$_{\rm AF}$ is shown in Fig. \ref{T1}.
%
%, it fits our data very well. The obtained doping dependence of $\theta$ is plotted in Fig. \ref{phasediagram}(b).
For $x_{\rm M}$~=~0.027, $^{75}$(1/$T_{\rm 1c}$)$_{\rm AF}$ is almost $T$-independent, which  intuitively demonstrates  that the system shows a quantum critical behavior. %$x_{\rm M}$ = 0.027 is a magnetic QCP.
The $T$-linear resistivity %observed  at doping concentration
supports this conclusion (see Fig. S12 \cite{supplemental_materials}).
We emphasize that $1/T_1$ and the resistivity are high-$T$ fingerprints of the magnetic QCP, while the peak of $\lambda _{\rm L}^{2}\left( 0 \right)$ is the direct evidence of QCP at the zero-$T$ limit.
In passing, we note that previous result on  BaFe$_2$(As$_{1-x}$P$_x$)$_2$ \cite{HashimotoScience} has created theoretical debates on its interpretation \cite{Chowdhury1,Nomoto,Levchenko,Chowdhury}.
Although several theoretical works showed that magnetic QCP can give rise to an enhanced   $\lambda _{\rm L}^{2}$ \cite{Nomoto,Levchenko,Chowdhury},
Chowdhury \emph{et. al.} warned that a phase separation could give rise to a decrease of superfluid density, thereby result in an increasing of $\lambda _{\rm L}^{2}\left( 0 \right)$\cite{Chowdhury1}. This was indeed the case in LaFeAsO$_{1-x}$F$_x$ system\cite{Luetkens}, where phase separation was evidenced by nuclear quadrupole resonance measurements\cite{Lang1}. %,Lang}.
However, as mentioned above, no indication of phase separation was seen in our samples by $^{23}$Na or $^{75}$As NMR spectra \cite{supplemental_materials,ZhouPRB}.
Our result therefore indicates that indeed a magnetic QCP can give rise to mass enhancement.

On the other hand, $x_{\rm c}$ = 0.032 is clearly far from $x_{\rm M}$ = 0.027, and thus the mass enhancement there  is $not$ related to the magnetic QCP. We note that  $T_{\rm s}$ extrapolates to zero around  $x_{\rm c}$ = 0.032 \cite{ZhouPRB}, at which the electrical resistivity also  shows a good $T$-linear behavior up to $T$ = 110 K ( see Fig. S12 \cite{supplemental_materials}). %found that we find that the nematic QCP should be
This result together with previous NMR \cite{ZhouPRB} and  Raman \cite{Blumberg} studies suggest that nematic fluctuations \cite{Schattner_1,Lederer_PNAS} exist  %around $x_{\rm c}$
above the superconducting dome.
 We conclude that the peak we observed at  $x_{\rm c}$=0.032 is an evidence   that a nematic QCP lies {\it beneath} the superconducting dome,
 %In coefficient $A$ in the resistivity fitting $\rho=\rho_0+AT^n$ shows two maximum in the $x$ dependence, which supports the existence of two distinct QCPs (see Supplemental Materials \cite{supplemental_materials}) .
 where the mass is enhanced by a factor $\sim$~2.5  due to a band renormalization caused by
 quantum nematic  fluctuations. %This is consistent with recent theories  \cite{Fitzpatrick,Lederer_PNAS}.
 It was theoretically shown by a Monte Carlo calculation that a nematic quantum fluctuations can lead to an enhancement of a factor $\sim$ 4 \cite{Lederer_PNAS}.
 %This conclusion is supported by the upper critical field that shows a maximal at $x$=0.032 (see Supplemental Materials).

%Although Raman scattering studies suggest that the nematic QCP is located at $x$ = 0.025\cite{Thorsm}, considering that Raman susceptibility couples to phonons, the position of nematic QCP is underestimated, which also means that it should be in the overdoped region\cite{Hinojosa}.
%Then our results will naturally suggest that a nematic QCP is inside superconducing dome at the
%Namely, the peak of $\lambda _{L}^{2}\left( 0 \right)$  at $x_c$ = 0.032 is due to a band renormalization associated with nematic quantum fluctuations.
%%In fact, recent theories have shown that the electrons can become massive around a nematic QCP, leading to non Fermi liquid behavior in the normal state as we observed by electrical transport measurement
%But the singularity of $\lambda _{L}^{2}\left( 0 \right)$ is not predicted by previous theoretical studies\cite{Chowdhury}, which might imply that some effects in iron-pnictides should be considered, such as orthorhombic strain due to lattice effect\cite{Paul_1}.
%Then our results might also give some clues about the origin of nematicity in iron-pnictides\cite{Fernades_np}. For example, by using a model of a spin driven nematic phase, elastic strain was found to suppress the fluctuations of the nematic order parameter itself and change the universality class of the nematic phase transition\cite{Fernandes_2,Karahasanovic}.

 %BaFe$_{2-x}$Co$_x$As or
 The existence of a nematic QCP seems to affect superconductivity of this system.
In BaFe$_2$(As$_{1-x}$P$_x$)$_2$ where $x_{\rm M}$ and $x_{\rm c}$ are too close or indistinguishable, there is a well-defined maximum in the doping dependence of $T_c$. In striking contrast,  $T_c$ of NaFe$_{1-x}$Co$_x$As  shows a weak  decrease  for $x\geq$0.027, %.has a very weak doping dependence in the region 0.02 $<$ $x$ $<$ 0.04,
 as seen in Fig. \ref{phasediagram} (b). This suggests that nematic fluctuations play a role to enhance the  pairing interaction. % for $x\geq$0.03.
 It is believed by many that superconductivity at low doping region is mediated by spin fluctuations with large momentum $q$. We speculate that  nematic fluctuation with   $q\sim$0  helps enhance pairing interaction as to prevent $T_c$ from a rapid decrease at high  $x$ beyond 0.027 where spin fluctuations are weakened. Same is probably true in the second dome of LaFeAsO$_{1-x}$F$_x$ (0.3 $<$ $x$ $<$ 0.8) where low energy fluctuations is weak but $T_c$ is higher than that in the first dome (0 $<$ $x$ $<$ 0.25)\cite{Yang}.
 % is increased beyond 0.03.
 %Some theoretical works  has already addressed this issue \cite{Kuo_SCI1,Lederer_PNAS}, and
 Our results provide strong motivations for further  investigations in this regards.
 %This is an important issue in a  broader context.   In the case of conventional $s$-wave superconductors,  phonons are the collective modes that mediate the interaction between two electrons. In the case of cuprate high-temperature superconductors and heavy-fermion superconductors,  spin fluctuations are likely the mediator \cite{Coleman_nat,pairing}. In principle, other collective modes can play the same role, thus our results provide strong motivations for further theoretical and experimental investigations on the nematic fluctuations as mediators. % in the case of iron-based superconductors.
Also, It would be a good future task to investigate how the pairing symmetry changes when nematic fluctuations are weakened at large $x$ where $T_{\rm c}$ decreases. % \cite{supplemental_materials}.

%It also suggests that at least spin and nematic fluctuations should be considered to fully comprehend the mechanism of superconductivity in Fe-based superconductors.
Meanwhile, in %the high-$T_c$  cuprate
YBa$_2$Cu$_3$O$_y$, quantum oscillation  shows that the effective mass is enhanced around the optimal doping \cite{Ramshaw}, where
%pseudogap line is actually an transition with
 the rotation symmetry was  found to be broken \cite{Daou,Zhao_1,Cuprate_namatic}, which suggests that there also exists a   QCP with nematic character. % close to the optimal doping  \cite{Nie}.
 Therefore, our results suggest a possible link between the two different classes of the high-$T_c$ superconductors %may also shed  lights on high-$T_c$ superconductivity in the cuprates
and will stimulate %will provoke
more studies on the  cuprates. % in this regard. %in the superconducting state .

In summary, we have systematically studied the zero-$T$-limit London penetration depth $\lambda_{\rm L}^2$(0) in NaFe$_{1-x}$Co$_x$As to diagnose the quantum critical behavior inside the superconducting dome. A nematic QCP is found inside the superconducting dome at $x_{\rm c}$ = 0.032, which is clearly distinguished from the magnetic QCP $x_{\rm M}$ = 0.027.
Our results indicate that the electron mass is enhanced near the nematic QCP due to band renormalization by nematic quantum fluctuations.

We thank S. A. Kivelson, J. Schmalian, S. Lederer,  D. H. Lee, Z.Q. Wang and S. Uchida for useful discussion. This work was partially supported by NSFC Grant No. 11634015, and MOST
 %the National Key R$\And$D Program
 of China (No. 2017YFA0302904 and No.2016YFA0300502).

\clearpage


\vspace{0.5cm}

\begin{references}

\bibitem{Cuprate_review}
P. A. Lee, N. Nagaosa, and X. G. Wen, Rev. Mod. Phys. \textbf{78}, 17 (2006).
	
\bibitem{Fernandes_review}
G. R. Stewart, Rev. Mod. Phys. \textbf{83}, 1589 (2011).

\bibitem{Cooper}  P. Gegenwart, Q. Si, and F. Steglich, Nat Phys \textbf{4}, 186 (2008).
%\bibitem{Cooper} Cooper, R. A. et. al. Anomalous criticality in the electrical resistivity ofLa$_{2-x}$Sr$_x$CuO$_4$. \emph{Science} \textbf{323}, 603¨C607 (2009).

\bibitem{HashimotoScience}
K. Hashimoto, K. Cho, T. Shibauchi, S. Kasahara, Y. Mizukami, R. Katsumata, Y. Tsuruhara, T. Terashima, H. Ikeda, M. A. Tanatar, H. Kitano, N. Salovich, R. W. Giannetta, P. Walmsley, A. Carrington, R. Prozorov, and Y. Matsuda, Science \textbf{336}, 1554 (2012).

%\bibitem{LSCO}Cooper, R. A.  \emph{et al.} %  Wang, Y., Vignolle, B., Lipscombe, O. J., Hayden, S. M., Tanabe, Y., Adachi, T., Koike, Y.,  Nohara, M., Takagi, H., Proust, C. $\And$ Hussey, N. E.
  %  Anomalous criticality in the electrical resistivity of La$_{2-x}$Sr$_x$CuO$_4$. \emph{Science} \textbf{323}, 603-607 (2009).
\bibitem{Chu2} J. H. Chu, H. H. Kuo, J. G. Analytis, and I. R. Fisher, Science \textbf{337}, 710 (2012).


\bibitem{ZhouNatCommun}
R. Zhou, Z. Li, J. Yang, D. L. Sun, C. T. Lin, and G. Q. Zheng, Nat. Commun. \textbf{4}, 2265 (2013).

\bibitem{Coleman_nat}P. Coleman and A. J. Schofield, Nature \textbf{433}, 226 (2005).

%\bibitem{Scalapino}
%D. J. Scalapino, Rev. Mod. Phys. \textbf{84}, 1383 (2012).



%\bibitem{Keimer_cuprate}B. Keimer, S. A. Kivelson,	M. R. Norman, S. Uchida	$\And$ J. Zaanen, From quantum matter to high-temperature superconductivity in copper oxides. \emph{Nature} \textbf{518}, 179-186 (2015).



\bibitem{pairing}
T. Moriya and K. Ueda, Adv. Phys. \textbf{49}, 555 (2000).

\bibitem{Xly}
E. P. Rosenthal, E. F. Andrade, C. J. Arguello, R. M. Fernandes, L. Y. Xing, X. C. Wang, C. Q. Jin, A. J. Millis, and A. N. Pasupathy, Nat. Phys. \textbf{10}, 225 (2014).

\bibitem{S4}
T. Iye, M. H. Julien, H. Mayaffre, M. Horvatic, C. Berthier, K. Ishida, H. Ikeda, S. Kasahara, T. Shibauchi, and Y. Matsuda, J. Phys. Soc. Jpn. \textbf{84}, 043705 (2015).

\bibitem{ZhouPRB}
R. Zhou, L. Y. Xing, X. C. Wang, C. Q. Jin, and G. Q. Zheng, Phys. Rev. B \textbf{93}, 060502 (2016).


\bibitem{Kuo_SCI1}
H. H. Kuo, J. H. Chu, J. C. Palmstrom, S. A. Kivelson, and I. R. Fisher, Science \textbf{352}, 958 (2016).

\bibitem{Fernandes_1}R. M. Fernandes and J. Schmalian, Phys. Rev. B \textbf{82}, 014521 (2010).

\bibitem{Kontani}
H. Kontani and S. Onari, Phys. Rev. Lett. \textbf{104}, 157001 (2010).



\bibitem{Chubukov_1}A. Chubukov and P. J. Hirschfeld, Phys. Today. \textbf{68}, 46 (2015).

\bibitem{Schmalian2014}
R. M. Fernandes, A. V. Chubukov, and J. Schmalian, Nat. Phys. \textbf{10}, 97 (2014).


\bibitem{Yang}
J. Yang, R. Zhou, L. L. Wei, H. X. Yang, J. Q. Li, Z. X. Zhao, and G. Q. Zheng, Chin. Phys. Lett. \textbf{32}, 107401 (2015).

%\bibitem{Si_iron}Q. Si, R. Yu $\And$ E. Abrahams, High-temperature superconductivity in iron pnictides and chalcogenides. \emph{Nature Reviews Materials} \textbf{1}, 16017 (2016).

\bibitem{Lederer_2}
S. Lederer, Y. Schattner, E. Berg, and S. A. Kivelson, Phys. Rev. Lett. \textbf{114}, 097001 (2015).

\bibitem{Lee}
Z.-X. Li, F. Wang, H. Yao, D.-H. Lee, Sci. Bull. \textbf{61}, 925 (2016).

\bibitem{Lederer_PNAS}
S. Lederer, Y. Schattner, E. Berg, and S. A. Kivelson, Proc. Natl. Acad. Sci. U.S.A. \textbf{114}, 4905 (2017).

\bibitem{Varma_1}C. M. Varma, K. Miyake, and S. Schmittrink, Phys. Rev. Lett. \textbf{57}, 626 (1986).

\bibitem{Ramshaw}
B. J. Ramshaw, S. E. Sebastian, R. D. McDonald, J. Day, B. S. Tan, Z. Zhu, J. B. Betts, R. X. Liang, D. A. Bonn, W. N. Hardy, and N. Harrison, Science \textbf{348}, 317 (2015).



\bibitem{Walmsley}
P. Walmsley, C. Putzke, L. Malone, I. Guillamon, D. Vignolles, C. Proust, S. Badoux, A. I. Coldea, M. D. Watson, S. Kasahara, Y. Mizukami, T. Shibauchi, Y. Matsuda, and A. Carrington, Phys. Rev. Lett. \textbf{110}, 257002 (2013).

%\bibitem{Prozorov_2}R. Prozorov $\And$ R. W. Giannetta, Magnetic penetration depth in unconventional superconductors. \emph{Supercond. Sci. Technol.} \textbf{19} R41-R67, (2006).


%\bibitem{IronbasedHosono}
%Y.~Kamihara, T.~Watanabe, M.~Hirano, and H.~Hosono.
%\newblock Iron-based layered superconductor LaO$_{1-x}$F$_{x}$FeAs(\emph{x}=0.05-0.12) with
%  $T_{c}$=26 K.
%\newblock {\em Journal of the American Chemical Society}, 130(11):3296--+,
%  2008.
%\bibitem{Stewart}
%G.~R. Stewart.
%\newblock Superconductivity in iron compounds.
%\newblock {\em Reviews of Modern Physics}, 83(4):1589--1652, 2011.


%\bibitem{transportQCP}
%S.~Kasahara, T.~Shibauchi, K.~Hashimoto, K.~Ikada, S.~Tonegawa, R.~Okazaki,
%  H.~Shishido, H.~Ikeda, H.~Takeya, K.~Hirata, T.~Terashima, and Y.~Matsuda.
%\newblock Evolution from non-fermi- to fermi-liquid transport via isovalent
%  doping in BaFe$_{2}$(As$_{1-x}$P$_{x}$)$_{2}$ superconductors.
%\newblock {\em Physical Review B}, 81(18), 2010.



%\bibitem{heatcapacityQCP}
%Z.~Walmsley, C.~Putzke, L.~Malone, I.~Guillamon, D.~Vignolles, C.~Proust,
 % S.~Badoux, A.~I. Coldea, M.~D. Watson, S.~Kasahara, Y.~Mizukami,
 % T.~Shibauchi, Y.~Matsuda, and A.~Carrington.
%\newblock Quasiparticle mass enhancement close to the quantum critical point in
%  BaFe$_{2}$(As$_{1-x}$P$_{x}$)$_{2}$.
%\newblock {\em Physical Review Letters}, 110(25), 2013.

%\bibitem{Stewart1}
%G.~R. Stewart.
%\newblock Non-fermi-liquid behavior in d- and f-electron metals.
%\newblock {\em Reviews of Modern Physics}, 73(4):797--855, 2001.


%\bibitem{SCwithoutphonons}
%P.~Monthoux, D.~Pines, and G.~G. Lonzarich.
%\newblock Superconductivity without phonons.
%\newblock {\em Nature}, 450(7173):1177--1183, 2007.

%\bibitem{Scalapino}
%D.~J. Scalapino.
%\newblock A common thread: The pairing interaction for unconventional
%  superconductors.
%\newblock {\em Reviews of Modern Physics}, 84(4):1383--1417, 2012.





\bibitem{NaFeAs}
D. R. Parker, M. J. Pitcher, P. J. Baker, I. Franke, T. Lancaster, S. J. Blundell, and S. J. Clarke, Chem. Commun. \textbf{16}, 2189 (2009).

%\bibitem{ChenXH}
%A. F. Wang, X. G. Luo, Y. J. Yan, J. J. Ying, Z. J. Xiang, G. J. Ye, P. Cheng, Z. Y. Li, W. J. Hu, and X. H. Chen, Phys. Rev. B \textbf{85}, 224521 (2012).



\bibitem{supplemental_materials} See Supplemental Material for additional data and analysis.




%\bibitem{NaFeAs_Dai}S. Li, C. de la Cruz, Q. Huang, G. F. Chen, T.-L. Xia, J. L. Lou, N. L. Wang, abd P. C. Dai, Structural and magnetic phase transitions in Na$_{1-\delta}$FeAs. Phys. Rev. B \textbf{80}, 020504 (2009).

%\bibitem{Kitagawa_NaFeAs}Kitagawa, K., Mezaki, Y., Matsubayashi, K., Uwatoko, Y. \& Takigawa, M. Crossover from Commensurate to Incommensurate Antiferromagnetism in Stoichiometric NaFeAs Revealed by Single-Crystal $^{23}$Na, $^{75}$As-NMR Experiments. \emph{J. Phys. Soc. Jpn.} \textbf{80}, 033705 (2011).



%\bibitem{Korringa}
%J.~Korringa.
%\newblock Nuclear magnetic relaxation and resonnance line shift in metals.
%\newblock {\em Physica}, 16(7-8):601--610, 1950.


%\bibitem{Tabuchi}Tabuchi, T. \emph{et al.} Evidence for a full energy gap in the nickel pnictide supercon-ductor LaNiAsO$_{1-x}$F$_x$ from $^{75}$As nuclear quadrupole resonance. \emph{Phys. Rev. B} \textbf{81}, 140509 (2010).

%\bibitem{Kitagawa}
%K.~Kitagawa, N.~Katayama, K.~Ohgushi, M.~Yoshida, and M.~Takigawa.
%\newblock Commensurate itinerant antiferromagnetism in BaFe$_{2}$As$_{2}$: $^{75}$As NMR
%  studies on a self-flux grown single crystal.
%\newblock {\em Journal of the Physical Society of Japan}, 77(11), 2008.

%\bibitem{ZhengPRL}
%G.-q. Zheng  {\it et al.}, Delocalized quasiparticles in the vortex state of an
%overdoped high-$T_c$ superconductor probed by $^{63}$Cu NMR. \emph{Phys. Rev. Lett.} {\bf 88},
%077003 (2002).




\bibitem{Brandt}
E. H. Brandt, Phys. Rev. B \textbf{37}, 2349 (1988).

%\bibitem{Tanaka}
%K. K. Tanaka, M. Ichioka, and S. Onari, Phys. Rev. B \textbf{93}, 094507 (2016).



\bibitem{Ghannadzadeh}
S. Ghannadzadeh, J. D. Wright, F. R. Foronda, S. J. Blundell, S. J. Clarke, and P. A. Goddard, Phys. Rev. B \textbf{89}, 054502 (2014).

\bibitem{Halperin}
S. Oh, A. M. Mounce, J. A. Lee, W. P. Halperin, C. L. Zhang, S. Carr, and P. Dai, Phys. Rev. B \textbf{87}, 174517 (2013).

\bibitem{Curro}
N. J. Curro, C. Milling, J. Haase, and C. P. Slichter,
Phys. Rev. B {\bf 62}, 3473 (2000).

\bibitem{Mitrovic}V. F. Mitrovi\'{c} , \emph{et. al.}, Spatially resolved electronic structure inside and outside the vortex cores of a high-temperature superconductor. Nature \textbf{413}, 501-504 (2001).

\bibitem{ZhengTl2201}
G.-q. Zheng, H. Ozaki, Y. Kitaoka, P.Kuhns, A. P. Reyes, and W. G. Moulton,
Phys. Rev. Lett. {\bf 88},  077003 (2002).


\bibitem{Kumagai}
K. Kakuyanagi, K. Kumagai, Y. Matsuda, and M. Hasegawa, Phys. Rev. Lett. \textbf{90}, 197003 (2003).

\bibitem{BrandtPRL}
E. H. Brandt, Phys. Rev. Lett. {\bf 66}, 3213 (1991).

\bibitem{Bi2212}
D. R. Harshman, E.H. Brandt, A.T. Fiory, M. Inui, D.B.Mitzi, L.F. Schneemeyer and J.V. Waszczak, Phys. Rev. B {\bf 47}, 2905 (1993).


%\bibitem{Oh}S. Oh, \emph{et. al.}, Spin pairing and penetration depth measurements from nuclear magnetic resonance in NaFe$_{0.975}$Co$_{0.025}$As, \emph{Phys. Rev. B} \textbf{87}, 174517 (2013).

%\bibitem{Shan}L. Shan, \emph{et. al.}, Observation of ordered vortices with Andreev bound states in Ba$_{0.6}$K$_{0.4}$Fe$_2$As$_2$, \emph{Nature Physics} \textbf{7}, 325-331 (2011).


%\bibitem{Brandt_2}E. H. Brandt, Magnetic-field variance in layered superconductors, \emph{Phys. Rev. Let.} \textbf{66}, 3213 (1991).

% \bibitem{Harshman}D. R. Harshman, \emph{et. al.}, Longitudinal disordering of vortex lattices in anisotropic superconductors, \emph{Phys. Rev. B} \textbf{47}, 2905 (1993).

% \bibitem{Kalisky}B. Kalisky, \emph{et. al.}, Behavior of vortices near twin boundaries in underdoped Ba(Fe$_{1-x}$Co$_x$)$_2$As$_2$, Phys. Rev. B 83, 064511 (2011).

%\bibitem{Prozorov_1}R. Prozorov, \emph{et. al.}, Anisotropic London penetration depth and superfluid density in single crystals of iron-based pnictide superconductors, \emph{Physica C} \textbf{469}, 582-589 (2009).
\bibitem{Gennes}
P. G. de Gennes, Superconductivity of Metals and Alloys (Westview Press, Oxford, UK, 1999).

\bibitem{Pardo}
E. Pardo, D.-X. Chen, and A. Sanchez, J. Appl. Phys. \textbf{96}, 5365 (2004)

\bibitem{Liu_1}
Z. H. Liu, P. Richard, K. Nakayama, G. F. Chen, S. Dong, J. B. He, D. M. Wang, T. L. Xia, K. Umezawa, T. Kawahara, S. Souma, T. Sato, T. Takahashi, T. Qian, Y. B. Huang, N. Xu, Y. B. Shi, H. Ding, and S. C. Wang, Phys. Rev. B \textbf{84}, 064519 (2011).

\bibitem{Ge}
Q. Q. Ge, Z. R. Ye, M. Xu, Y. Zhang, J. Jiang, B. P. Xie, Y. Song, C. L. Zhang, P. C. Dai, and D. L. Feng, Phys. Rev. X \textbf{3}, 011020 (2013).


\bibitem{Parker_XRD}
D. R. Parker, M. J. P. Smith, T. Lancaster, A. J. Steele, I. Franke, P. J. Baker, F. L. Pratt, M. J. Pitcher, S. J. Blundell, and S. J. Clarke, Phys. Rev. Lett. \textbf{104}, 057007 (2010).

\bibitem{Okada}
 T. Okada, H. Takahashi, Y. Imai, K. Kitagawa, K. Matsubayashi, Y. Uwatoko, and A. Maeda, Physica C: Superconductivity \textbf{494}, 109 (2013).


\bibitem{Chowdhury1}
D. Chowdhury, J. Orenstein, S. Sachdev, and T. Senthil, Phys. Rev. B \textbf{92}, 081113 (2015).

%\bibitem{Gordon}R. T. Gordon, \emph{et. al.}, Doping evolution of the absolute value of the London penetration depth and superfluid density in single crystals of Ba(Fe$_{1-x}$Co$_x$)$_2$As$_2$. Phys. Rev. B \textbf{82}, 054507 (2010).








%\bibitem{Moon_1}E. G. Moon $\And$ Subir Sachdev, Quantum critical point shifts under superconductivity: Pnictides and cuprates, \emph{Phys. Rev. B} \textbf{82}, 104516 (2010).

%\bibitem{Chang}J. Chang, \emph{et. al.}, Magnetic field controlled charge density wave coupling in underdoped YBa$_2$Cu$_3$O$_{6+x}$. \emph{Nature Communications} \textbf{7}, 11494 (2016).



\bibitem{Nomoto}
T. Nomoto and H. Ikeda, Phys. Rev. Lett. \textbf{111}, 167001 (2013).

\bibitem{Levchenko}
A. Levchenko, M. G. Vavilov, M. Khodas, and A. V. Chubukov, Phys. Rev. Lett. \textbf{110}, 177003 (2013).


\bibitem{Chowdhury}
D. Chowdhury, B. Swingle, E. Berg, and S. Sachdev, Phys. Rev. Lett. \textbf{111}, 157004 (2013).

\bibitem{Co-NMR-prl_Ning}
F. L. Ning, K. Ahilan, T. Imai, A. S. Sefat, M. A. McGuire, B. C. Sales, D. Mandrus, P. Cheng, B. Shen, and H. H. Wen, Phys. Rev. Lett. \textbf{104}, 037001 (2010).

\bibitem{itinerantelectronmagnetism}
T. Moriya, J. Mag. Mag. Mat \textbf{100}, 261 (1991).


\bibitem{Korringa}
J. Korringa, Physica \textbf{16}, 601-610 (1950).

\bibitem{Luetkens}
H. Luetkens, H.-H. Klauss, M. Kraken, F. J. Litterst, T. Dellmann, R. Klingeler, C. Hess, R. Khasanov, A. Amato, C. Baines, M. Kosmala, O. J. Schumann, M. Braden, J. Hamann-Borrero, N. Leps, A. Kondrat, G. Behr, J. Werner and B. B\"{u}chner, Nat. Mater. \textbf{8}, 305¨C309 (2009).

%\bibitem{Lang1}
%G. Lang, H.-J. Grafe, D. Paar, F. Hammerath, K. Manthey, G. Behr, J. Werner, and B. B\"{u}chner, Phys. Rev. Lett. 104, 097001 (2010).


\bibitem{Lang1}
G. Lang, L. Veyrat, U. Gr\"{a}fe, F. Hammerath, D. Paar, G. Behr, S. Wurmehl, and H.-J. Grafe, Phys. Rev. B \textbf{94}, 014514 (2016);

G. Lang, H.-J. Grafe, D. Paar, F. Hammerath, K. Manthey, G. Behr, J. Werner, and B. B\"{u}chner, Phys. Rev. Lett. 104, 097001 (2010).

%\bibitem{Co-NMR-prl_Ning}Ning, F. L. \emph{et al.} % Ahilan, K.,  Imai, T., Sefat, A. S., McGuire, M. A., Sales, B. C.,  Mandrus, D., Cheng, P., Shen, B.  $\And$ Wen, H. H.
%Contrasting spin dynamics between underdoped and overdoped Ba(Fe$_{1-x}$Co$_x$)$_2$As$_2$. \emph{Phys. Rev. Lett.} \textbf{104}, 037001 (2010).

%\bibitem{itinerantelectronmagnetism}
%T.~Moriya.
%\newblock Theory of itinerant electron magnetism.
%\newblock {\em Journal of Magnetism and Magnetic Materials}, 100(1-3):261--271,
%1991.



%\bibitem{Thorsm}V. K. Thorsm{\o}le, \emph{et. al.}, Critical quadrupole fluctuations and collective modes in iron pnictide superconductors. \emph{Phys. Rev. B} \textbf{93}, 054515 (2016).

%\bibitem{Luan}L. Luan, \emph{et. al.}, Local Measurement of the Superfluid Density in the Pnictide Superconductor Ba(Fe$_{1-x}$Co$_x$)$_2$As$_2$ across the Superconducting Dome. Phys. Rev. Lett. \textbf{106}, 067001 (2011).

%\bibitem{Hinojosa}A. Hinojosa, J. Cai, $\And$ A. V. Chubukov, Raman resonance in iron-based superconductors: The magnetic scenario. \emph{Phys. Rev. B} \textbf{93}, 075106 (2016).
%\bibitem{Galeis}
%Gallais, Y. et al. Observation of Incipient Charge Nematicity in BaFe$_{2-x}$Co$_x$As$_2$. \emph{Phys. Rev. Lett.} \textbf{111}, 267001 (2013).

%\bibitem{Fitzpatrick}Fitzpatrick, A. L., Kachru, S., Kaplan, J. \& Raghu, S. Non-Fermi-liquid fixed point in a Wilsonian theory of quantum critical metals. \emph{Phys. Rev. B} \textbf{88}, 125116 (2013).


%\bibitem{Lederer_1}S. Lederer, Y. Schattner, E. Berg $\And$ S. A. Kivelson, Superconductivity and bad metal behavior near a nematic quantum critical point. \emph{arxiv:}1612.01542 (2016).

%\bibitem{Lederer_unpublished}
%S. Lederer,  Y. Schattnerb, E. Berg, and S. A. Kivelson, Superconductivity and non-Fermi liquid behavior near
%a nematic quantum critical point, Proc Natl
%Acad Sci USA {\bf 114},4905 (2017).




%\bibitem{Paul_1}I. Paul $\And$ M. Garst, Lattice effects on nematic quantum criticality in metals. \emph{arXiv:}1610.06168v1 (2016).

%\bibitem{Fernades_np}Fernandes, R. M., Chubukov, A. V. \& Schmalian, J. What drives nematic order in iron-based superconductors? \emph{Nat. Phys.} \textbf{10}, 97-104 (2014).

%\bibitem{Fernandes_2}R. M. Fernandes, A. V. Chubukov, J. Knolle, I. Eremin, and J. Schmalian, Preemptive nematic order, pseudogap, and orbital order in the iron pnictides. \emph{Phys. Rev. B} \textbf{85}, 024534 (2012).

%\bibitem{Karahasanovic}U. Karahasanovic $\And$ J. Schmalian, Elastic coupling and spin-driven nematicity in iron-based superconductors. \emph{Phys. Rev. B} \textbf{93}, 064520 (2016).

%\bibitem{Yamase}H. Yamase $\And$ R. Zeyher, Superconductivity from orbital nematic fluctuations. \emph{Phys. Rev. B} \textbf{88}, 180502(R) (2013).

\bibitem{Blumberg}
V. K. Thorsmolle, M. Khodas, Z. P. Yin, C. L. Zhang, S. V. Carr, P. C. Dai, and G. Blumberg, Phys. Rev. B \textbf{93}, 054515 (2016).

\bibitem{Schattner_1}
Y. Schattner, S. Lederer, S. A. Kivelson, and E. Berg, Phys. Rev. X \textbf{6}, 031028 (2016).

%\bibitem{Kontani}
%Kontani, H. $\And$  Onari, S.,
% Orbital-Fluctuation-Mediated Superconductivity in Iron Pnictides: Analysis of the Five-Orbital Hubbard-Holstein Model.
%\emph{Phys. Rev. Lett.} {\bf 104}, 157001 (2010).

%\bibitem{Xia}J. Xia, \emph{et. al.}, Polar Kerr-Effect Measurements of the High-Temperature YBa$_2$Cu$_3$O$_{6+x}$ Superconductor: Evidence for Broken Symmetry near the Pseudogap Temperature. \emph{Phys. Rev. Lett.} \textbf{100}, 127002 (2008).

 \bibitem{Daou}
 R. Daou, J. Chang, D. LeBoeuf, O. Cyr-Choiniere, F. Laliberte, N. Doiron-Leyraud, B. J. Ramshaw, R. X. Liang, D. A. Bonn, W. N. Hardy, and L. Taillefer, Nature \textbf{463}, 519 (2010).

 \bibitem{Cuprate_namatic}
 Y. Sato, S. Kasahara, H. Murayama, Y. Kasahara, E. G. Moon, T. Nishizaki, T. Loew, J. Porras, B. Keimer, T. Shibauchi, and Y. Matsuda, Nat. Phys. \textbf{13}, 1074 (2017).


\bibitem{Zhao_1}
L. Zhao, C. A. Belvin, R. Liang, D. A. Bonn, W. N. Hardy, N. P. Armitage, and D. Hsieh, Nat. Phys. \textbf{13}, 250 (2017).

%\bibitem{Nie}L. Nie, G. Tarjus, S. A. Kivelson, \emph{Proc. Natl. Acad. Sci. U.S.A.} \textbf{111}, 7980-7985 (2014).


%\bibi.tem{Fujita}K. Fujita \emph{et. al.}, \emph{Science} \textbf{344}, 612-616 (2014).


%\bibitem{Metlitski}M. A. Metlitski, D. F. Mross, S. Sachdev, $\And$ T. Senthil, Cooper pairing in non-Fermi liquids. \emph{Phys. Rev. B} \textbf{91}, 115111 (2015).

%\bibitem{Li_1}Z.-X. Li, F. Wang, H. Yao $\And$ D.-H. Lee, The nature of effective interaction in cuprate superconductors: a sign-problem-free quantum Monte-Carlo study. \emph{arXiv:}1512.04541 (2015).

%\bibitem{Cavanagh}D. C. Cavanagh, A. C. Jacko, and B. J. Powell, Phys. Rev. B \textbf{92}, 195138 (2015).

%\bibitem{WangAF} Wang, A.F. et al, A crossover in the phase diagram of NaFe$_{1-x}$Co$_x$As determined by electronic transport, \emph{New J. Phys.} {\bf 15}, 043048 (2015).

\end{references}
\end{document}